\documentstyle[twocolumn,aps,prl,overcite]{revtex} 
\input epsf

\tighten
\begin{document}
 


\title{Quintessence, Cosmic Coincidence, and the Cosmological
Constant }

\author{ Ivaylo Zlatev,$^1$ Limin Wang$^1$
and Paul J. Steinhardt$^{1,2}$
}

\address{$^1$Department of Physics and Astronomy,
University of Pennsylvania,
Philadelphia, PA 19104 
 \vspace{.1in} \\
$^2$Department of Physics, Princeton University, Princeton, NJ 08540
}

\maketitle

\begin{abstract}

Recent observations suggest that a large fraction
of the energy density of the universe  has negative
pressure.
One explanation is vacuum energy density;
 another   is   quintessence in the form of
  a  scalar field slowly evolving down a potential.
 In either case, a key problem is to explain why the energy density
 nearly coincides with the matter density today. The densities
 decrease at different rates as the universe expands, so coincidence
 today 
 appears to require that their ratio  
  be set to a specific, infinitessimal value in the early
 universe.
  In this paper, we introduce the
  notion of a ``tracker field," a form of quintessence,  and
  show how it may explain the coincidence, adding new
  motivation for the quintessence scenario.

\end{abstract} 

A number of recent observations suggest that 
$\Omega_m$, the ratio of the (baryonic plus dark) matter density
to the critical density, is significantly less than unity\cite{ostriker}.  
Either the universe is open, or there is some additional energy 
density $\rho$ sufficient to reach $\Omega_{total} =1$,
 as predicted by inflation.  Measurements of
the cosmic microwave background, 
the mass power spectrum\cite{ostriker,Turner,Wang98b}, and,
most explicitly, the luminosity-red shift relation observed for
Type Ia supernovae\cite{supernova}, all suggest that the missing energy should
possess  negative pressure ($p$) and   equation-of-state
($w \equiv p/\rho$).  
One candidate for the
missing energy is vacuum energy density  or
cosmological constant, $\Lambda$ for which $w=-1$. 
The resulting cosmological model, $\Lambda$CDM, consists of a mixture of vacuum 
energy and cold dark matter. 
Another possibility  is QCDM cosmologies based on a mixture of 
cold dark matter and quintessence ($-1<w\le 0$), a
slowly-varying, spatially inhomogeneous component.\cite{Cald98}
An example of quintessence
is the energy associated with a scalar field ($Q$) slowly evolving
down its potential $V(Q)$.\cite{Ratra,Friem,Cald98,Ferreira}  
Slow evolution is needed 
to obtain negative pressure,
$p=\frac{1}{2}\dot{Q}^2 -V(Q)$, so that the
kinetic energy density is less than the potential energy density.

Two difficulties arise from all of these scenarios. The first is
the fine-tuning problem: Why is the
missing energy density today
 so small compared to typical particle physics scales?
If $\Omega_m \sim 0.3$ today
the missing energy
density is of order 
$10^{-47}$~GeV$^4$, which appears to require the introduction of 
new mass scale 14 or so orders of magnitude smaller than the 
electroweak scale.
A second difficulty is the ``cosmic coincidence"
problem:\cite{CMBdata}   
Since the missing energy density
and the matter density
decrease at  different rates as the universe expands,
it appears that their ratio must be set to a specific, infinitessimal
value in the very early universe in order for the two densities to nearly
coincide today, some 15 billion years later.

What seems most ideal is a model in which
the energy density in the $Q$-component is comparable to the radiation
density (to within a few order of magnitude) at the end of inflation, say.
If there were some rough equipartition of energy following reheating
among several thousands of degrees of freedom,
one might  expect the energy density of the $Q$-component
 to be two or so orders of magnitude smaller
than the total radiation density.
One would want that the energy density of the $Q$-component somehow
tracks below the background density for most of the
history of the universe, and, then, only recently, grows to dominate the
energy density  and drive it into a period of 
accelerated expansion.  
The models we present  will
do all this and more even though there is  only one adjustable parameter.
The models are extremely insensitive to initial conditions --- 
variations in the 
initial ratio of the $Q$-energy density to the matter density 
by nearly 100 orders of magnitude do not affect  the cosmic history.
 The models are
in excellent agreement with current measurements of the cosmic 
microwave background, large scale structure, and cosmic acceleration.
We also find that the models predict a relation between $\Omega_m$
and the acceleration of the universe. These properties suggest
a new perspective for the quintessence models, perhaps placing them
on equal footing with the more conventional $\Lambda$ models.

The models considered in this Letter are 
based on the notion of ``tracker fields,"
a form of quintessence in which the tracker field $Q$ rolls down a 
potential $V(Q)$ according to an attractor-like solution to the 
equations-of-motion.  
The tracker solution is an attractor in the sense that a very wide range of 
initial conditions for $Q$ and $\dot{Q}$ rapidly approach a
common evolutionary track, so that the cosmology is insensitive to the
initial conditions.   
Tracking has an advantage similar to 
inflation in that a wide range of initial conditions
is funneled into the same final condition.  
This contrasts with most quintessence potentials studied previously
in the literature\cite{Friem,Cald98} which require very fine adjustment of
the initial value of $Q$ (as well as parameters in the potential) to obtain a 
suitable cosmology.
We introduce the term ``tracker" because there is a subtle but important
difference from attractor solutions in dynamical systems.
Unlike a standard attractor,
the tracker solution is not 
a fixed point (in the sense of a fixed point solution of a system of
autonomous differential equations of motion \cite{Ferreira}): 
the ratio of the $Q$-energy to the background matter or
radiation density  changes steadily as $Q$ proceeds down its track.  
This is desirable because one is interested in having the $Q$-energy
ultimately overtake the background density and drive the universe
towards an accelerating phase. 
This contrasts with the  ``self-adjusting" solutions 
recently discussed by Ferreira and Joyce\cite{Ferreira} based on
$V(Q)=M^4 exp(\beta Q)$ potentials.   Self-adjusting solutions are 
more nearly
true attractors in that
$\Omega_Q$ remains constant for a constant background equation-of-state 
($\Omega_Q$ changes slightly when the universe transforms from radiation- to
matter-domination).
This means, for example, 
that $\Omega_Q$ is constant throughout the matter-dominated
epoch.  For constant $\Omega_Q$, 
satisfying the constraints from structure formation  requires that
$\Omega_Q$  be less than 0.2 and $\Omega_m$  exceed 0.8, which
runs into conflict with current best 
estimates\cite{ostriker,Wang98b,Cald98} of $\Omega_m$ and 
produces a decelerating universe in conflict with recent supernovae 
results.\cite{supernova} 
So, the interesting and significant features of
 tracking are that: (a) as for the self-adjusting case, a wide range 
 of initial conditions are drawn towards a common cosmic history; but, (b)
the tracking solutions do not ``self-adjust" to the 
background equation-of-state, but, instead,
maintain some finite difference in the equation-of-state
such that the $Q$-energy  ultimately
dominates and the universe enters a period of acceleration. 
Compared to the self-adjusting case, tracking does not require any
additional parameters and allows a much wider range of potentials.

Tracker solutions exist for a very wide 
class of potentials,\cite{newtrack} including
potentials in which 
$d\, {\rm ln V}/dQ$ is slowly decreasing  
as $Q$ rolls downhill.
(The self-adjusting potentials
correspond to constant $d\, {\rm ln V}/dQ$.)
 The energy density of the tracker field decreases
as $1/a^{3(1+w_Q)}$ where $w_Q$ remains constant 
or varies slowly
in each epoch of 
the universe but changes 
sharply
when the background expansion of
the  universe changes from 
radiation- to matter- to quintessence-dominated.
The value of $w_Q$ differs from the background equation-of-state such that
the value of $\Omega_Q$ increases as the universe ages and, for most 
potentials, increases more rapidly as the universe ages.  Hence, 
it is more likely that $\Omega_Q$ grows to order unity late in the history
in the universe compared to earlier.

We will consider two examples: $V(Q)= M^{(4+\alpha )} Q^{-\alpha}$ 
and $V(Q)= M^4 \,[ {\rm exp} (M_p/Q)-1]$,
where $M$ is the one free parameter 
 and $M_p$ is the Planck mass. 
 For any given $V$, there is a family
 of tracker solutions parameterized by $M$.  The value of $M$
 is fixed by the measured value of $\Omega_m$.
The potentials are suggested by
particle physics models with
dynamical symmetry breaking or nonperturbative 
effects,\cite{AC,barreiro,gaillard,barrow,partphys}
although we consider it
 premature to justify our concept at this formative stage
on the basis of fundamental physics. Our purpose, rather, is to show
that a simply-parameterized fluid with the desired properties is
physically possible.
Pioneering studies of the inverse
power-law case  have been done by
Ratra and Peebles.\cite{Ratra} 
Here we point out some 
  additional important properties and generalizations,  and, then,
explain how all of these properties  are relevant to quintessence and
the coincidence problem and possibly the fine-tuning problem.

The tracker field $Q$ satisfies the equation-of-motion:
\begin{equation} \label{eom}
\ddot{Q} + 3 H \dot{Q}+ V'(Q) = 0
\end{equation}
where $V'(Q)$ is the derivative of $V$ with respect to $Q$ and
$H$ is the Hubble parameter.
For the inverse
power-law potential,  $Q$ 
has a tracker solution\cite{Ratra}  which 
maintains the condition:
\begin{equation} \label{tt}
V'' = (9/2)(1-w_Q^2)((\alpha +1)/\alpha)  H^2.
\end{equation}
The condition that $\rho_Q$ is beginning to dominate today
means that $Q$ must be  ${\cal O}( M_p)$ today
since $V'' \approx \rho_Q/Q^2$ and $H^2 \approx \rho_Q/M_p^2$. 

 The one free parameter, $M$, is determined by the 
observational constraint, $\Omega_Q \approx 0.7$ today.  
Here is where the fine-tuning 
issue must be considered.
 To have $\Omega_Q
\approx 0.7$ today
requires $V(Q \approx M_p) \approx \rho_m$, where $\rho_m \approx
10^{-47}$~GeV$^4$ is the current matter density;  this imposes the 
constraint
$M \approx (\rho_m M_p^{\alpha})^{1/(\alpha+4)}$.   For low values of $\alpha$
or for the exponential potential, this forces $M$ to be a tiny 
mass as low as 1~meV for the exponential case.  However, we note
that   $M > 1$~GeV ---  comparable to 
particle physics scales ---   is possible for  $\alpha \gtrsim 2$. Hence, 
while this is not our real aim, it is interesting to note that
the tracker
solution  
does not  require
the introduction of a  new mass hierarchy in fundamental parameters.

To address the coincidence problem 
--- removing the need to  tune initial 
conditions in order for the matter and missing energy densities to nearly coincide today ---
our proposal relies on the 
tracking behavior  of $Q$ in a background of standard cosmology.
Let us first consider $V(Q)= M^{(4+\alpha)} Q^{-\alpha}$ for $\alpha
\ge 1$.
For any fixed $M$, the tracker solution is determined by  Eq.~(\ref{tt}),
We shall call the energy density in the $Q$-field as a function
of $z$ along the tracker solution    $\bar{\rho}_Q(z)$.    
If initial conditions are set at $z=z_i$, at the end of inflation, say,
then one possibility is that the initial energy density in $Q$
is less than the attractor value, $\rho_Q (z_i)< \bar{\rho}_{Q}(z_i)$.
In this case, the field remains frozen until $H^2$
decreases to the point where   Eq.~(\ref{tt}) is satisfied.
See Figure 1.
After that point,
 $Q$ begins rolling down the potential,  maintaining the
condition in Eq.~(\ref{tt})
 as it rolls along. A second possibility is
that the initial energy density in $Q$ is greater than the tracker
value but less than the background radiation density,
 $\bar{\rho}_Q(z_i) <\rho_Q(z_i) < 
\rho_B(z_i)$. This includes the case of equipartition after
reheating.  In this case  $Q$ starts rolling
 down the potential immediately and so rapidly that its 
kinetic energy $\frac{1}{2} \dot{Q}^2$
dominates over the potential energy density $V(Q)$. The kinetic energy 
density 
red shifts as $1/a^6$ and eventually $Q$ comes  nearly to a stop
at $Q \approx 0.5 (\rho_Q(z_i)/\rho_B(z_i))^{1/2} M_p$. By this point, 
$Q$ has  
fallen   below
the tracker solution, $\bar{\rho}_Q$.  Now, $Q$ remains nearly frozen
and $H$
decreases
  until  Eq.~(\ref{tt}) is satisfied.
Then, $Q$ tracks the same solution  as before.
Hence, any initial $\rho_Q$ less than the initial background
radiation density, including equipartition initial conditions, 
leads to the same tracker solution and the same cosmology.

The only troublescome case is if $Q$ dominates over the
background radiation density initially,  $\rho_Q \gg \rho_B$.
In this case case, 
$Q$ grows to a value greater than $M_p$  before it slows down; this
 overshoots the tracker solution to such an 
extent that the tracker  is not reached by the present
epoch and $\rho_Q$ is insignificant today.  
On the other hand,  the initial condition $\rho_Q \gg \rho_B$ 
seems unlikely.

\begin{figure}[t]
 \epsfxsize=3.3 in \epsfbox{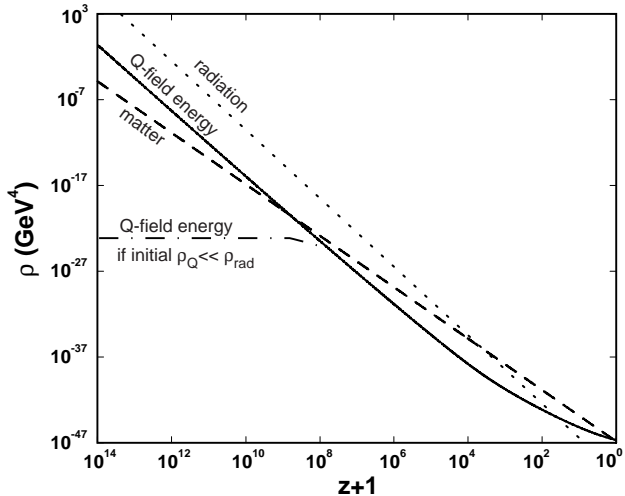}
\caption{
The evolution of the energy densities 
 for  a quintessence component
with $V(Q)= M^4 \,  [{\rm exp} (M_p/Q)-1]$ potential.
The solid line is 
 where 
  $\rho_Q$ is initially
comparable to the radiation density and immediately
evolves according to tracker solution.  The dot-dashed curve is if,
for some reason, $\rho_Q$ begins at a much smaller value. The
field is frozen and $\rho_Q$  is constant until the dot-dashed
curve runs into tracker solution,
leading to the same cosmology today: $\Omega_m=0.4$ and $w_Q=-0.65$.
}
\end{figure}

For the pure inverse power-law potential, the energy density
decays as a constant power of the scale factor $a$;
{\it i.e.}, $\rho_Q \propto a^{-3(1+w_Q)}$  and
\begin{equation} \label{wq}
w_Q \approx \frac{\frac{\alpha}{2} w_B -1}{1 + \frac{\alpha}{2}},
\end{equation} 
where this approximation is valid
so long as  $\rho_B \gg \rho_Q$.
The variable  $w_B$ is the
equation-of-state of the background: $w_B=0$ in the matter-dominated
epoch and $w_B=1/3$ in the radiation-dominated epoch. That is, 
the $Q$-component acts as a fluid with constant equation-of-state, 
but its value of $w_Q$ depends both on its effective potential
$V(Q)$ and on the background.  The effect of the background is through
the $3H\dot{Q}$ in the equation-of-motion for $Q$, Eq.~(\ref{eom});
when $w_B$ changes, $H$ also changes which, in turn, changes 
rate at which the tracker field evolves down the potential.

The second remarkable feature of the tracker solutions is
that  $w_Q$ automatically decreases to a negative value as the universe
transforms from radiation- to matter-dominated, whether $w_Q$ is 
positive ($\alpha >6$) or negative ($\alpha <6$) 
in the radiation-dominated epoch. 
This means that  $\rho_Q$ decreases at a slower rate than the matter density.
 Consequently, the matter-dominated era 
cannot last forever. Eventually,  perhaps close to the
present epoch, the $Q$-component overtakes the matter density.

The third remarkable feature is that, once the $Q$-component
begins to dominate, its behavior 
changes again: the $Q$-field slows to nearly a
stop causing  the equation-of-state $w_Q$ to decrease towards -1.
Hence, the universe begins a period of accelerated expansion.

If 
$\Omega_m \ge 0.2$ today, then the $Q$-component has dominated for
only a short time and $w_Q$ has not had time to reach -1 today.
For $\alpha \gg 1$, $w_Q$ is nearly 1/3 during the radiation 
epoch, nearly zero  in the matter dominated epoch, and
has fallen to a value   $\gtrsim -1/3$  today. The predicted 
current value is larger than recent supernovae results 
suggest.\cite{garnavich98}
As  $\alpha $ is made smaller,  $w_Q$ is  smaller at each stage along
the tracker solution, including today.
For $\alpha\le 6$, for example, 
we find  $w_Q \gtrsim -0.8$ for $\Omega_m \ge 0.2$,
 in closer accord with recent supernovae results.\cite{garnavich98} 
\begin{figure} [t]
\epsfxsize=3.3 in \epsfbox{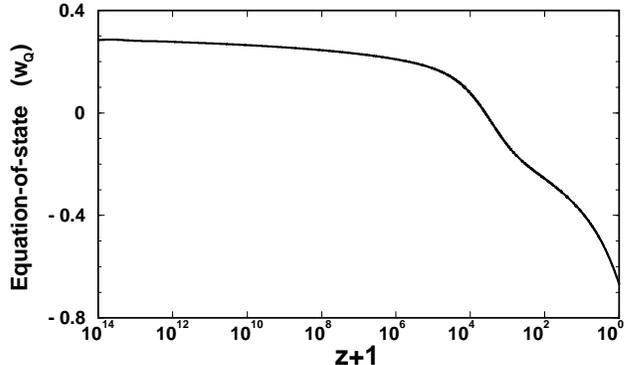}
\caption{$w_Q$ vs. $z$ for the model in Figure 1.
 During the radiation-dominated epoch (large $z$),
$w_Q \approx 1/3$ and the $Q$-energy density tracks the radiation 
background. During the matter-dominated epoch, $w_Q$ becomes somewhat
negative  (dipping down to $w_Q \approx -0.2$ beginning at $z=10^4$)
until $\rho_Q$ overtakes the matter density; then,
$w_Q$ plummets towards -1 and the universe begins to accelerate.
}
\end{figure}

The exponential potential,
$V(Q) = M^4 \, [{\rm exp} (M_p/Q)-1]$,  is an example
of combining inverse power-law models,
which introduces yet another generic feature of tracking.
The exponential potential
can be expanded in inverse powers of $Q$, where
 the dominant power $\alpha$ varies from high values to 
low values as $Q$ evolves towards larger values, causing $w_Q$ to
decrease as the universe ages.
As a result, $\Omega_Q$ grows more rapidly as the universe ages, making
it more likely that $\Omega_Q$ dominates later in the history of the universe
rather than earlier.
We use this model
for the purposes of illustration.
In Figure 1 we show  the evolution of $\rho_Q$ relative
to the matter and radiation density. We show  the case
where $\rho_Q$ is comparable to the radiation density at the 
end of inflation (solid curve) and also the case where $\rho_Q$
is initially much smaller. The latter case produces precisely the
same cosmology once the $Q$-field starts rolling.
In Figures 2-4, we illustrate the evolution
of $w_Q$,  the comparison of the linear mass power
spectrum to recent Automatic Plate Measuring (APM) large-scale structure survey results,\cite{Peac97}
and the cosmic microwave background  temperature anisotropy
power spectrum compared to recent data from the COBE, Big Plate and CAT
experiments.\cite{CMBdata}

\begin{figure} [t]
 \epsfxsize=3.3 in \epsfbox{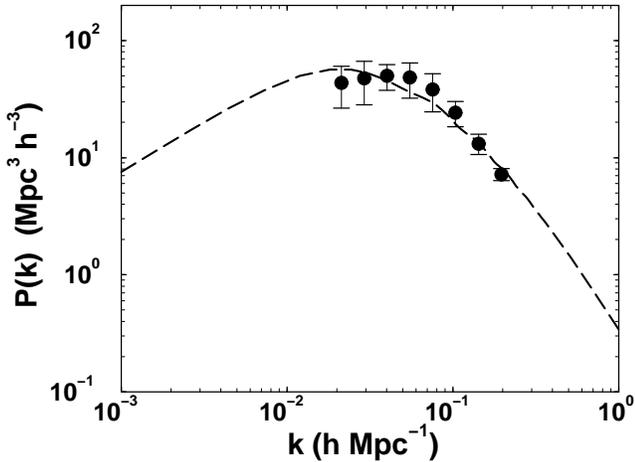}
\caption{
The linear mass power spectrum for the model in Figure 1
assuming Hubble parameter $H_0=65$~km/sec/Mpc, compared
to the Automatic Plate Measuring (APM) galaxy survey.  
}
\end{figure}

An important prediction to emerge from the tracker field models is a 
relation between $\Omega_m$ and $w_Q$ today (for fixed $h$).  
For any given potential,
the prediction is precise: fixing $\Omega_m$ today also fixes the
one free parameter, $M$.  Consequently, $w_Q$ is determined, as well.
Even without restricting to a particular potential,
the trend is clear:  smaller $\Omega_m$ means that the tracker field 
has been dominating longer and $w_Q$ is closer to -1 today. 
Given that $\Omega_m
\ge 0.2$, we have found that it is not 
 possible  to obtain $w_Q <-0.8$ without adding artificial 
complications to the potential. The bound is very weakly $h$-dependent.
This value is
significantly different from $w=-1$ for a cosmological constant and
one can hope to detect this difference. 

One brief word should be added about the future of the universe:
as $Q$ continues to evolve, it slows down and $w_Q$ approaches 
arbitrarily close to -1.  So, the universe expands as if there is a 
fixed non-zero cosmological constant, even though the reality is that $Q$ is
slowly oozing its way downhill.

We thank R. Caldwell and A. Liddle  for  useful conversations.
This research was supported by the US Department of Energy grants
 DE-FG02-95ER40893 (Penn) and
DE-FG02-91ER40671 (Princeton).
We have modified the CMBFAST  software
routines\cite{cmbfast} for our computations.

\begin{figure}[t]
 \epsfxsize=3.3 in \epsfbox{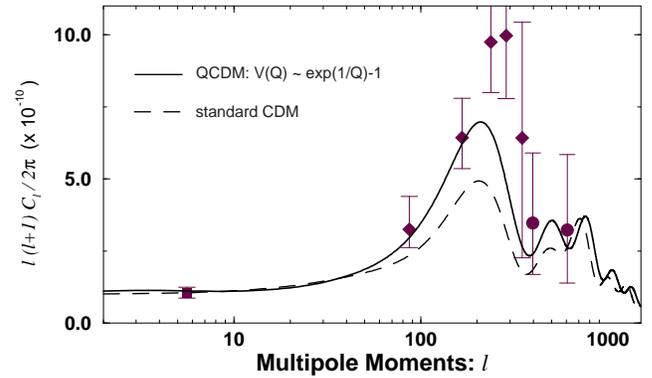}
\caption{
The cosmic microwave background anisotropy power spectrum for
the model in Figure 1 compared to the standard cold dark matter
model and recent data.\protect\cite{CMBdata}
}
\end{figure}



\end{document}